\documentclass[12pt]{article}
\usepackage[utf8]{inputenc}
\usepackage[english]{babel}
\usepackage{amsmath, amsfonts,amssymb}
\usepackage{graphicx}
\usepackage{float}
\usepackage{geometry}
\usepackage{hyperref}
\usepackage{physics}
\usepackage{subcaption}
\usepackage{url}

\hypersetup{
	bookmarksopen=true,
	colorlinks=true,
	linkcolor=blue,
	citecolor=blue,
	urlcolor=black,	
	linktoc=all,
	pdftitle={Stiff string},
	pdfauthor={N. Guarin-Zapata},
	pdfkeywords={Wave equation, Strings, Bending stiffness, Guitars, Fretboard},
	pdfsubject={Class project},
	pdfpagemode=UseOutlines,
	pdfstartview=FitH
}

\author{Nicolás Guarín Zapata\\ nguarinz@eafit.edu.co}
\title{\textbf{Design of a fretboard using the stiff string equation}}

\begin{document}
\maketitle
\setlength{\parskip}{1cm}

\abstract{\footnotesize Guitar fretboards are designed based on the equation of
the ideal string. That is, it neglecs several factors as nonlinearities and
bending stiffness of the strings. Due to this fact, intonation of guitars along
the whole neck is not perfect, and guitars have right tuning just in an
\emph{average} sense. There are commercially available fretboards that differ
from the tradictional design.\footnote{One example is the \cite{patent} by the
Company True Temperament AB, where each fretboard is made using CNC processes.}
As a final application of this work we would like to redesign the fretboard
layout considering the effects of bending stiffness.

The main goal of this project is to analyze the differences between the
differences in the solution for vibrations of the ideal string and a stiff
string. These differences should lead to changes in the fret distribution for a
guitar, and, hopefully improve the overall intonation of the instrument. We will
start analyzing the ideal string equation and after a good understanding of this
analytical solution we will proceed with the, more complex, stiff equation.
Topics like separation of variables, Fourier transforms,  and Perturbation
analysis might prove useful during the course of this project.}

\section{Modeling}
The ideal string equation is a second order partial differential equation, while
the stiff equation includes an extra term that turns it into a fourth order
partial differential equation. This additional term takes into account the
bending stiffness that is normally neglected in the ideal string.\footnote{This
term is probably more important for electric guitars, since they use steel
strings. And also for bass guitars, since they have thicker strings.}

\begin{figure}[h]
\centering
\includegraphics[height=6cm]{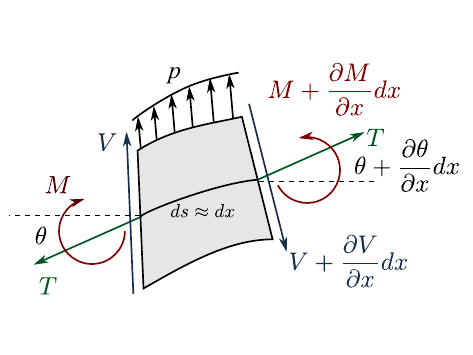}
\caption{Forces diagram over an element of the string with length $dx$. }
\label{fig:string_forces}
\end{figure}

Let's take a small segment with small displacements in the vertical direction
$w$. The body diagram is shown in Figure  \ref{fig:string_forces}, where we
assumed that the tension is constant along the  length element.

Summing forces in the normal to the centerline of the element yields
\[T\sin\left(\theta + \pdv{\theta}{x}\dd{x}\right) - T\sin\theta + V
- \left(V + \pdv{V}{x}\dd{x}\right) + p\dd{x} = (\rho A \dd{x})\pdv[2]{w}{t} \, .\]

If we assume that the angles are small, then $\sin\theta \approx \theta$, thus
\[T\left(\theta + \pdv{\theta}{x}\dd{x}\right) - T\sin\theta + V
- \left(V + \pdv{V}{x}\dd{x}\right) + p\dd{x} = (\rho A \dd{x})\pdv[2]{w}{t} \, ,\]
expanding and dividing by $\dd{x}$
\[T\pdv{\theta}{x} - \pdv{V}{x} + p = \rho A \pdv[2]{w}{t}\, .\]

We know that $\tan\theta = \pdv{w}{x}$, then $\sec^2\theta \pdv{\theta}{x}
= \pdv[2]{w}{x}$, and since we considered small angles $\sec^2\theta \approx 1$,
replacing this in the equation we obtain
\begin{equation}
T\pdv[2]{w}{t} - \pdv{V}{x} + p = \rho A \pdv[2]{w}{t} \, .
\label{eq:force_balance}
\end{equation}

Summing moments about the center of the element and omitting higher order
differentials, yields
\[M + \left(M + \pdv{M}{x}\dd{x}\right) - V\dd{x} = 0 \, ,\]
that  after expansion reads
\[V = \pdv{M}{x}\, .\]

We can use the same assumptions that are done in the Euler-Bernoulli beams
modeling, i.e., that the bending moment is proportional to the linearized
curvature
\[M = EI\pdv[2]{w}{x}\, ,\]
that leads to
\[\pdv{V}{x} = \pdv[2]{M}{x} = \pdv[2]{x}\left(EI \pdv[2]{w}{x}\right)\, ,\]
and replacing in \eqref{eq:force_balance}, yields
\begin{equation}
 T\pdv[2]{w}{x} - \pdv[2]{x}\left(EI \pdv[2]{w}{x}\right) + p = \mu \pdv[2]{w}{t}\, ,
 \label{eq:motion_equation}
\end{equation}
with $\mu =\rho A$ the linear mass density. If we assume that the bending
stiffness $EI$ is constant, and consider that there are no body forces, we
obtain
\begin{equation}
 T\pdv[2]{w}{x} - EI \pdv[4]{w}{x} = \mu \pdv[2]{w}{t}\, .
 \label{eq:stiff_string_orig}
\end{equation}

\subsection{Nondimensional form}
If we start from equation \ref{eq:stiff_string_orig} we can rewrite the equation
in non-dimensional form as
\begin{equation}
  \pdv[2]{u}{\xi} - \epsilon \pdv[4]{u}{\xi} - \alpha^2 \pdv[2]{u}{\tau} = 0\, ,
  \label{eq:stiff_string}
\end{equation}
with $u = w/L$, $\xi = x/L$, $\tau = \omega t$, $\alpha^2 = L^2 \mu \omega^2/T
= L^2 \omega^2/c^2$, $\epsilon = EI/(L^2 T)$, $L$ the length of the string,
$\omega$ a characteristic frequency of the system, $c$ the phase speed for the
ideal string. When, $EI$ is small the equation can be rewritten as
\begin{equation}
  \pdv[2]{u}{\xi} - \alpha^2 \pdv[2]{u}{\tau} = 0\, ,
  \label{eq:ideal_string}
\end{equation}
meaning that we are neglecting the bending stiffness of the string.

\subsection{Solution of the PDE}
If we take the Fourier transform of equation \ref{eq:stiff_string} we obtain
\begin{equation}
  \pdv[2]{U}{\xi} - \epsilon \pdv[4]{U}{\xi} - \kappa^2 U = 0\, ,
  \label{eq:stiff_string_freq}
\end{equation}
with $u = w/L$, $\xi = x/L$, $\tau = \omega t$, $\kappa^2 = L^2 \mu \omega^2/T
= L^2 \omega^2/c^2$, $\epsilon = EI/(L^2 T)$, $L$ the length of the string,
$\omega$ a characteristic frequency of the system, $c$ the phase speed for the
ideal string, and $U$ is the Fourier transform of $u$. We know that the solution
for the time part of the PDE is of the form
\(A \sin(\omega t) + B\cos(\omega t)\), but we are more interested in the
spatial part. The reason for this interest is to find the eigenvalues of the
differential equation, that, ultimately, leads to the eigenfrequencies.

The solution of the resulting differential equation is
\begin{align*}
u(s) =& C_{1} \cos{\left (\frac{\sqrt{2} s}{2 \sqrt{\epsilon}} \sqrt{\alpha - 1} \right )}
  + C_{2} \sin{\left (\frac{\sqrt{2} s}{2 \sqrt{\epsilon}} \sqrt{\alpha - 1} \right )} \\
 &+ C_{3} \sinh{\left (\frac{\sqrt{2} s}{2 \sqrt{\epsilon}} \sqrt{\alpha + 1} \right )}
  + C_{4} \cosh{\left (\frac{\sqrt{2} s}{2 \sqrt{\epsilon}} \sqrt{\alpha + 1} \right )} \enspace ,
\end{align*}
with $\alpha=\sqrt{1 + 4\epsilon\kappa^2}$. With boundary conditions
\begin{align*}
u(0) = 0,\quad &\dv{u}{s} (0)=0\\
u(1) = 0,\quad &\dv{u}{s} (1)=0 \enspace ,
\end{align*}
If we solve for $C_3$ and $C_4$ the first two equations we find
\[C_4 = -C_1\quad C_3 = -\frac{C_2\sqrt{\alpha - 1}}{\sqrt{\alpha + 1}} \enspace\]
giving
\begin{align*}
u(s) = &C_{1} \cos{\left (\frac{\sqrt{2} s}{2 \sqrt{\epsilon}} \sqrt{\alpha - 1} \right )}
  - C_{1} \cosh{\left (\frac{\sqrt{2} s}{2 \sqrt{\epsilon}} \sqrt{\alpha + 1} \right )} - \\
  &\frac{C_{2} \sqrt{\alpha - 1}}{\sqrt{\alpha + 1}} \sinh{\left (\frac{\sqrt{2} s}{2 \sqrt{\epsilon}} \sqrt{\alpha + 1} \right )} + C_{2} \sin{\left (\frac{\sqrt{2} s}{2 \sqrt{\epsilon}} \sqrt{\alpha - 1} \right )} \enspace .
\end{align*}

The other boundary conditions lead to the system of equations
\[\begin{bmatrix}
A_{11} &A_{12}\\
A_{21} &A_{22}
\end{bmatrix}
\begin{Bmatrix}C_1\\ C_2 \end{Bmatrix} = 0 \enspace ,\]
with
\begin{align*}
A_{11} &= \cos{\left (\frac{\sqrt{2} \sqrt{\alpha - 1}}{2 \sqrt{\epsilon}} \right )}- \cosh{\left (\frac{\sqrt{2} \sqrt{\alpha + 1}}{2 \sqrt{\epsilon}} \right )}\\
A_{12} &= - \frac{\sqrt{\alpha - 1}}{\sqrt{\alpha + 1}} \sinh{\left (\frac{\sqrt{2} \sqrt{\alpha + 1}}{2 \sqrt{\epsilon}} \right )} + \sin{\left (\frac{\sqrt{2} \sqrt{\alpha - 1}}{2 \sqrt{\epsilon}} \right )}\\
A_{21} &= - \frac{\sqrt{2} \sqrt{\alpha - 1}}{2 \sqrt{\epsilon}} \sin{\left (\frac{\sqrt{2} \sqrt{\alpha - 1}}{2 \sqrt{\epsilon}} \right )} - \frac{\sqrt{2} \sqrt{\alpha + 1}}{2 \sqrt{\epsilon}} \sinh{\left (\frac{\sqrt{2} \sqrt{\alpha + 1}}{2 \sqrt{\epsilon}} \right )}\\
A_{22} &= \frac{\sqrt{2} \sqrt{\alpha - 1}}{2 \sqrt{\epsilon}} \cos{\left (\frac{\sqrt{2} \sqrt{\alpha - 1}}{2 \sqrt{\epsilon}} \right )} - \frac{\sqrt{2} \sqrt{\alpha - 1}}{2 \sqrt{\epsilon}} \cosh{\left (\frac{\sqrt{2} \sqrt{\alpha + 1}}{2 \sqrt{\epsilon}} \right )}
\end{align*}

To obtain solutions that are nontrivial, we need to make $\det(A)=0$, i.e.,
\begin{align*}
&\sqrt{\alpha - 1} \cos{\left (\frac{\sqrt{2} \sqrt{\alpha - 1}}{2 \sqrt{\epsilon}} \right )} \cosh{\left (\frac{\sqrt{2} \sqrt{\alpha + 1}}{2 \sqrt{\epsilon}} \right )} - \sqrt{\alpha - 1} \\
 &-\frac{1}{\sqrt{\alpha + 1}} \sin{\left (\frac{\sqrt{2} \sqrt{\alpha - 1}}{2 \sqrt{\epsilon}} \right )} \sinh{\left (\frac{\sqrt{2} \sqrt{\alpha + 1}}{2 \sqrt{\epsilon}} \right )} = 0 \enspace ,
\end{align*}
or
\begin{align}
 &\sqrt{\alpha - 1} \cos{\left (\frac{\sqrt{2} \sqrt{\alpha - 1}}{2 \sqrt{\epsilon}} \right )} - \frac{\sqrt{\alpha - 1}}{\cosh{\left (\frac{\sqrt{2} \sqrt{\alpha + 1}}{2 \sqrt{\epsilon}} \right )}} - \nonumber\\
 &\frac{1}{\sqrt{\alpha + 1}} \sin{\left (\frac{\sqrt{2} \sqrt{\alpha - 1}}{2 \sqrt{\epsilon}} \right )} \tanh{\left (\frac{\sqrt{2} \sqrt{\alpha + 1}}{2 \sqrt{\epsilon}} \right )} = 0
 \label{eq:characteristic}
\end{align}
This is the characteristic equation of our problem. we need to solve this
equation numerically to find the roots.

\subsection{Perturbation solution}
We can solve equation \ref{eq:characteristic} for the eigenvalues using
numerical methods. From a design point of view, it would be easier to have some
analytic expressions that could be used to compute the desired parameters. Thus,
we can use a perturbation method to find an approximated solution to this
problem \cite{book:perturbation}, and try to obtain some analytic expressions.

Let us assume solutions of the form
\begin{align*}
&u(s) = u_0(s) + \epsilon u_1(s) + \epsilon^2 u_2(s) + \cdots\\
&\kappa^2 = \kappa_0^2 + \epsilon \kappa_1^2 + \epsilon^2  \kappa_2^2 + \cdots
\end{align*}

If we substitute these in the differential equations and group by powers of
$\epsilon$, keeping the first two powers, we obtain
\begin{align*}
  &\left[\pdv[2]{s} + \kappa_0^2\right]u_0 = 0\\
  &\left[\pdv[2]{s} + \kappa_0^2\right]u_1 = \pdv[4]{u_0}{s} - \kappa_1^2 u_0 \enspace .
\end{align*}

The first equation is satisfied by the pairs
\[\kappa_0 = n \pi,\quad u_0(x) = C_1 \sin(n\pi s) \enspace ,\]
and for the second equation to have non-singular solution we get that it needs
to be orthogonal to the first equation,
\[\int_{0}^{1} u_0\left[\pdv[4]{s} - \kappa_1^2\right]u_1\dd{s} = 0 \enspace ,\]
i.e.,
\[\kappa_1 = n^2 \pi^2 \, .\]

If we solve the second differential equation we find that $u_1(s)=C_3 \sin(n\pi s)
+ C_4 \cos(n\pi s)$, and applying boundary conditions we find that $C_4=0$. Thus
\begin{align*}
&u \approx (C_1 + C_3\epsilon) \sin(n\pi s)\\
&\kappa^2 \approx n^2\pi^2 + \epsilon n^4 \pi^4 \, ,
\end{align*}
or
\begin{subequations}
  \begin{align}
  &u \approx C_1 \sin(n\pi s)\\
  &\kappa^2 \approx n^2\pi^2 + \epsilon n^4 \pi^4 \, , \label{eq:pert_eigs}
  \end{align}
  \label{eq:pert_solution}
\end{subequations}
where we absorbed the constants into a single one.

\begin{figure}[H]
\centering
	\begin{subfigure}[b]{0.45\textwidth}
		\includegraphics[width=\textwidth]{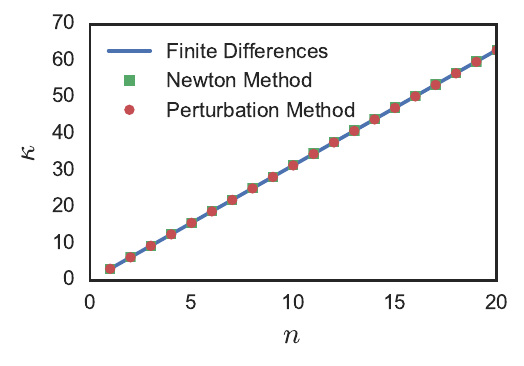}
		\caption{$\epsilon = 10^{-6}$.}
	\end{subfigure}
	\begin{subfigure}[b]{0.45\textwidth}
		\includegraphics[width=\textwidth]{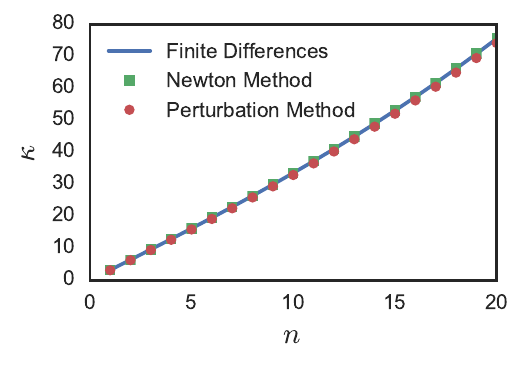}
		\caption{$\epsilon = 10^{-4}$.}
	\end{subfigure}\\
	\begin{subfigure}[b]{0.45\textwidth}
		\includegraphics[width=\textwidth]{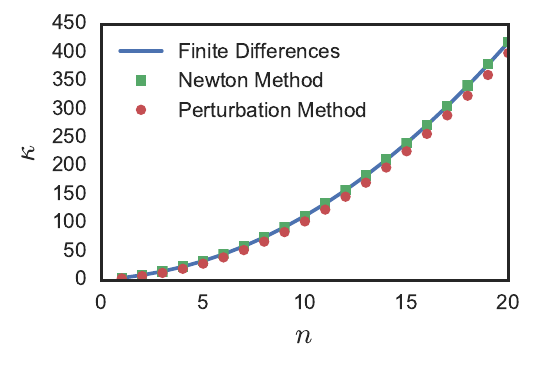}
		\caption{$\epsilon = 10^{-2}$.}
	\end{subfigure}
\caption{Eigenvalues for the stiff string for different values of $\epsilon$.}
\label{fig:stiff_string}
\end{figure}

This approximation is valid for small $\epsilon$.  Figure \ref{fig:stiff_string}
presents this approximation compared with the solution with Newton method using
these values as initial points. This figure also shows a solution using Finite
Differences with 1001 points. We can see that the approximation are good for
small \(\epsilon\) values.

Let us compute its value for a real example. A steel G string in an electric
guitar ($196$ Hz) has a perturbation parameter of
\[\epsilon \equiv \frac{E I}{TL^2} =  \frac{\pi E r^4}{4 TL^2} = \frac{\pi (2 \times 10^{11} \mbox{ Pa}) (0.4046 \times 10^{-3} \mbox{ m})^4}{4 (65.508 \mbox{ N})(0.6477 \mbox{ m})^2} \approx 9.725 \times 10^{-6} \, ,\]
that gives a small correction for the eigenvalues. What tells us that is a good
approximation to consider the string as an ideal string rather than one with
bending stiffness.

\section{Fretboard layouts}
This section describes the fretboard layout based on the eigenvalues obtained
from the ideal and stiff equations. For that we need to chose the temperament
for the instrument. That is, we need to pick the frequencies relationships
between consecutive notes. We are interested in a guitar with equal temperament
\cite{wiki:temperament}, i.e., that the ratio between two consecutive notes is
constant. And the octaves are made of 13 notes. This let us with a ratio of
$r = \sqrt[12]{2}\approx 1.0595$.

\subsection{Ideal string}
Based on the ideal string equation \cite{wiki:string_vibration,
book:vibration_cont}. We know that the fundamental frequency is given by
\begin{equation}
f_0 = \frac{v}{2 L_0}\, ,
\end{equation}
being $v=\sqrt{T/\mu}$ the wave speed, $T$ is the tension on the string, $\mu$
is the linear mass density, and $L_0$ is the string length. Then, the frequency
for the $n$th fret is given by
\[r^n f_0 = \frac{v}{2 L_n}\, , \]
where $L_n$ is the length of the part of the string that vibrates.

Solving for $L_n$, we get
\[L_n = \frac{v}{2 r^n f_0} = \frac{v}{2 r^n \left(\frac{v}{2L_0}\right)} = \frac{L_0}{r^n} \, ,\]
that is, the length of the vibrating part of the string is inversely
proportional to the increase in frequency.

The \emph{fret layout} refers to the distance from the nut, then we want the
difference between the overall distance and the vibrating part
\begin{equation}
  x_n = L_0 - L_n = L_0\left(1 - \frac{1}{r^n}\right)\, .
  \label{eq:fret_layout}
\end{equation}

\begin{figure}[h]
  \centering
  \includegraphics[width=6in]{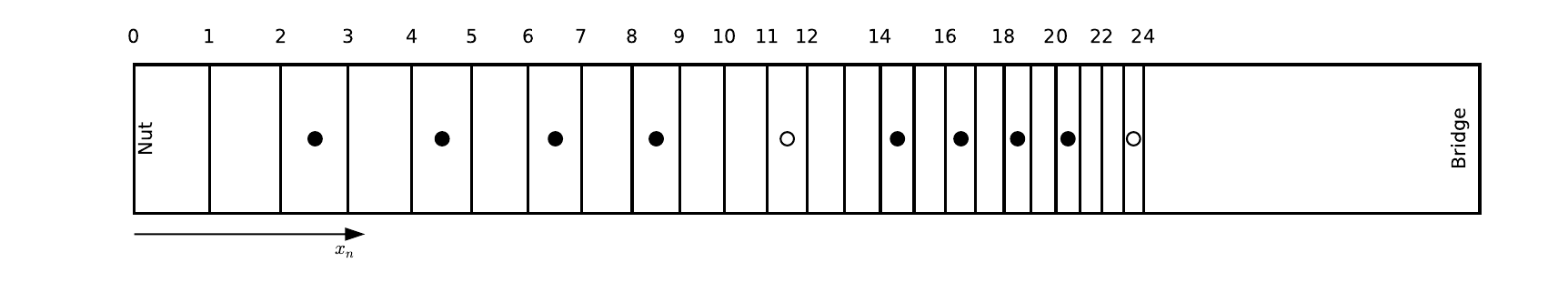} 
  \caption{Fret layout for an instrument with 24 frets.}
  \label{fig:fret_layout}
\end{figure}

And we can see that the distribution of frets depends solely in the ratio
between the length of the string and the \emph{vibrating length}. Figure
\ref{fig:fret_layout} presents a depiction of this distribution for the frets.

\subsection{Stiff string}
Based on equation \ref{eq:pert_eigs} we can conclude that the frequency of
oscillation of a stiff string is given by
\[f_0 = \frac{1}{DL}\sqrt{\frac{T}{\pi \rho}} \sqrt{1 + \pi^2 \epsilon_0} = \frac{1}{DL}\sqrt{\frac{T}{\pi \rho}} \sqrt{1 + \frac{\pi^3 E D^4}{64 L_0^2 T}}\, \]
with \(\epsilon = \frac{\pi E D^4}{64 TL_0^2}\). We want a string with
\emph{vibrating length} \(L_n\) and frequency \(r^n f_0\). This leads to the
equation
\[r^n f_0 = \frac{1}{DL}\sqrt{\frac{T}{\pi \rho}} \sqrt{1 + \frac{\pi^3 E D^4}{64 L_n^2 T}}\, ,\]
or
\begin{equation}
\frac{r^n}{L_0^2}[1 + \pi^2\epsilon_0] = \frac{1}{L_n^2}\left[1 + \pi^2\epsilon_0\frac{L_0^2}{L_n^2}\right]\, .
\label{eq:lengths}
\end{equation}

Solving equation \ref{eq:lengths} for \(L_0^2\) we obtain
\[\frac{L_n^2}{L_0^2} = \frac{1 \pm \sqrt{1 + 4 r^{2n} \gamma (1 + \gamma)}}{2\gamma^{2n}(1 + \gamma)}\, ,\]
with \(\gamma = \pi^2 \epsilon_0\). Only the solution with plus sign is of
interest since it has as limit case the ideal string result when
\(\gamma \rightarrow 0\). Thus
\[L_n = \frac{L_0\left[1 + \sqrt{1 + 4r^{2n}\gamma(1 + \gamma)}\right]^{1/2}}{r^n \sqrt{2 (1 + \gamma)}}\, ,\]
and
\begin{equation}
x_n = L_0 - L_n = L_0\left(1 - \frac{\left[1 + \sqrt{1 + 4 r^{2n}\gamma(1 + \gamma)}\right]^{1/2}}{r^n \sqrt{2 (1 + \gamma)}}\right)\, .
\label{eq:fret_layout_stiff}
\end{equation}

Although the expression for the fret layout has been cast in a similar fashion
than equation \ref{eq:fret_layout}, it should be noted that the parameter
\(\gamma\) depends on both, geometric and material parameters. Particularly,
\(\gamma\) is a function of \(L_0\) itself.

Since this new fret layout depends on properties of the material and the length
scale \(L_0\), we need to consider a particular set of strings. We now focus our
attention in the strings ESXL110 \cite{daddario_ESXL110} from D'Addario, the
diameters and tensions are presented in table \ref{tab:strings}. We also
consider a Young modulus of steel (\(E=200\) GPa), this is not completely right,
since the thicker strings are not made of a single material, but have a core
made of one material and are wounded with another material that mostly add mass
but not bending stiffness.
\begin{table}[h]
\centering
\begin{tabular}{cccc}
\hline 
\textbf{String} & \textbf{Note} & \textbf{Diameter (mm)} & \textbf{Tension (N)} \\ 
\hline 
1 & E & 0.2540 & 72.128 \\ 
2 & B & 0.3302 & 68.404 \\ 
3 & G & 0.4318 & 73.696 \\ 
4 & D & 0.6604 & 81.732 \\ 
5 & A & 0.9144 & 84.672 \\ 
6 & E & 1.1684 & 75.166 \\ 
\hline 
\end{tabular} 
\caption{Diameter and tensions for D'Addario Nickel wound string ESXL110
\cite{daddario_ESXL110}.}
\label{tab:strings}
\end{table}

\begin{figure}[h]
  \centering
  \includegraphics[width=4 in]{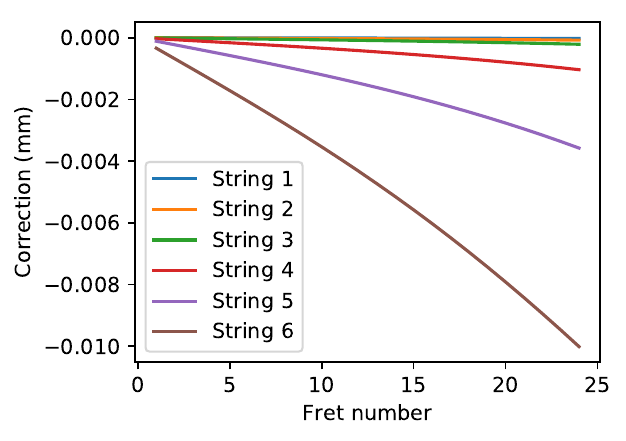} 
  \caption{Correction for fret positioning for string of length scale \(L_0 = 635\) mm.}
  \label{fig:fret_corrections}
\end{figure}

The corrections needed for each string are presented in Figure
\ref{fig:fret_corrections}. We considered a length scale of 635 mm (25 in).

Based on these corrections we depicted the the new fret layout that is presented
in Figure \ref{eq:fret_layout_stiff}, as expected, the larger corrections appear
for thicker strings.
\begin{figure}[h]
  \centering
  \includegraphics[width=6in]{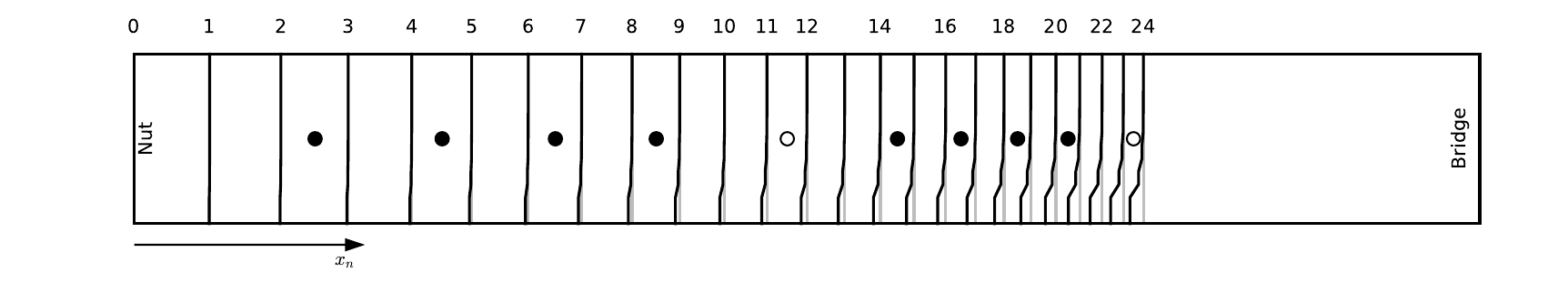} 
  \caption{Fret layout for an instrument with 24 frets and strings with length
  scale \(L_0 = 635\) mm.}
  \label{fig:fret_layout_stiff}
\end{figure}

\section{Conclusions}
We developed a model for the design of fretted instruments layout that considers
the bending stiffness of strings. Higher accuracy  can be achieved used
numerical methods but the use of analytical ones allows to write explicit
formula for the fret positioning.

Geometric (length scale and diameter), material (Young modulus), and loading
(tension) parameters appear explicitly in the expression for fret positioning.
Surprisingly, the mass density of the material does not appear in it.

As expected the strings with larger corrections are the thicker ones. This
conclusion might be misleading, since thicker strings are commonly made of an
inner core of steel (in electric guitars) or nylon (in classical guitars) and
have a wounding to add mass (and then lower the pitch). These strings can be
viewed as composite strings, and the model developed in the present document
does not cover this case. Fletcher presents a method to modify the equations to
include these ffects in reference \cite{fletcher1964}.

There are different causes for inharmonicities such as large amplitudes in the
motion of the strings that lead to non-linear responses, and  changes in tension
along the string, and they are not considered. Nevertheless, these effects have
been considered in different studies in the past \cite{shankland1939,
keller1959, young1952}.


\begin{thebibliography}{9}
\bibitem{daddario_ESXL110}
D'Addario \& Company, Inc. D'Addario String: XL Nickel Wound. Retrieved May 26,
2017, from
\url{http://www.daddario.com/DADProductDetail.Page?ActiveID=3769&productid=15&productname=ESXL110_Nickel_Wound__Regular_Light__Double_BallEnd__10_46}

\bibitem{book:instruments_physics}
Fletcher, N. H., \& Rossing, T. (2012). The physics of musical instruments.
Springer Science \& Business Media.

\bibitem{book:guitar_tech}
French, R. M. (2012). Technology of the Guitar. Springer Science \& Business Media.

\bibitem{fletcher1964}
Fletcher, H. (1964). Normal vibration frequencies of a stiff piano string. The
Journal of the Acoustical Society of America, 36(1), 203-209.

\bibitem{book:perturbation}
Holmes, Mark H. Introduction to perturbation methods (2012). Vol. 20. Springer
Science \& Business Media.

\bibitem{keller1959}
Keller, J. B. (1959). Large amplitude motion of a string. American Journal of
Physics, 27(8), 584-586.

\bibitem{book:vibration_cont}
Arthur W. Leissa, Mohamad S. Qatu (2011). Vibration of Continuous Systems.
McGraw-Hill.

\bibitem{wiki:string_vibration}
String vibration. (2017). In Wikipedia, The Free Encyclopedia. Retrieved April
12, 2017, from \url{https://en.wikipedia.org/wiki/String_vibration}

\bibitem{wiki:temperament}
Equal temperament. (2017). In Wikipedia, The Free Encyclopedia. Retrieved April
12, 2017, from \url{https://en.wikipedia.org/wiki/Equal_temperament}

\bibitem{shankland1939}
Shankland, R. S., \& Coltman, J. W. (1939). The departure of the overtones of a
vibrating wire from a true harmonic series. The Journal of the Acoustical
Society of America, 10(3), 161-166.

\bibitem{patent}
Thidell, A. (2010). U.S. Patent No. 7,728,210. Washington, DC: U.S. Patent and
Trademark Office.

\bibitem{yong2006}
Yong, D. (2006). Strings, chains, and ropes. SIAM review, 48(4), 771-781.

\bibitem{young1952}
Young, R. W. (1952). Inharmonicity of plain wire piano strings. The Journal of
the Acoustical Society of America, 24(3), 267-273.

\end{thebibliography}
\end{document}